\documentclass[aps,notitlepage,nofootinbib,twocolumn,longbibliography,10pt]{revtex4-1}

\newcommand{\beq}{\begin{eqnarray}}
\newcommand{\eeq}{\end{eqnarray}}
\newcommand{\beqq}{\begin{eqnarray*}}
\newcommand{\eeqq}{\end{eqnarray*}}

\newcommand{\be}{\begin{equation}}
\newcommand{\ee}{\end{equation}}
\newcommand{\barr}{\begin{array}}
\newcommand{\earr}{\end{array}}

\newcommand{\sign}{\text{sign}}

\newcommand{\vf}{v_F}

\newcommand{\ve}{{\varepsilon}}
\newcommand{\vef}{{\varepsilon_F}}

\usepackage[colorlinks=true,citecolor=blue,linkcolor=blue]{hyperref}
\usepackage{graphicx}
\usepackage{dcolumn}
\usepackage{bm}
\usepackage{color}
\usepackage{tabularx}
\usepackage{comment}
\usepackage{float}
\usepackage{amsmath}
\usepackage{gensymb}
\usepackage{mathtools}

\begin{document}

\begin{titlepage}

\widetext

\title{Large enhancement of thermopower at low magnetic field in compensated semimetals}

\author{Xiaozhou Feng}
\author{Brian Skinner}
\affiliation{Department of Physics, The Ohio State University, Columbus,
Ohio 43202, USA}

\setcounter{equation}{0}
\setcounter{figure}{0}
\setcounter{table}{0}

\makeatletter
\renewcommand{\theequation}{S\arabic{equation}}
\renewcommand{\thefigure}{S\arabic{figure}}
\renewcommand{\thetable}{S\Roman{table}}
\renewcommand{\bibnumfmt}[1]{[S#1]}
\renewcommand{\citenumfont}[1]{S#1}

\date{\today}

\begin{abstract}
The thermoelectric properties of conductors with low electron density can be altered significantly by an applied magnetic field. For example, recent work has shown that Dirac/Weyl semimetals with a single pocket of carriers can exhibit a large enhancement of thermopower when subjected to a sufficiently large field that the system reaches the extreme quantum limit, in which only a single Landau level is occupied. Here we study the magnetothermoelectric properties of \emph{compensated} semimetals, for which pockets of electron- and hole-type carriers coexist at the Fermi level. We show that, when the compensation is nearly complete, such systems exhibit a huge enhancement of thermopower starting at a much smaller magnetic field, such that $\omega_c \tau > 1$, and the stringent conditions associated with the extreme quantum limit are not necessary. We discuss our results in light of recent measurements on the compensated Weyl semimetal tantalum phosphide, in which an enormous magnetothermoelectric effect was observed. We also calculate the Nernst coefficient of compensated semimetals, and show that it exhibits a maximum value with increasing magnetic field that is much larger than in the single band case. In the dissipationless limit, where the Hall angle is large, the thermoelectric response can be described in terms of quantum Hall edge states, and we use this description to generalize previous results to the multi-band case.

\end{abstract}


\maketitle

\draft

\vspace{2mm}

\end{titlepage}

\section{Introduction}
\label{sec:Intro}

The thermoelectric effect is the generation of an electrical voltage difference $\Delta V$ from a temperature difference $\Delta T$ applied across a material. The thermoelectric effect has been an important topic in physics for over a hundred years, since it allows one to convert waste heat into useful electrical power \cite{ioffe_semiconductor_1957, shakouri_recent_2011}. The magnitude of the thermoelectric effect is quantified by the thermopower, or Seebeck coefficient, which can be defined as $S_{xx} = -\Delta V / (\Delta T)$, where both $\Delta V$ and $\Delta T$ are both measured along the same direction $x$ and in conditions where no current is flowing. Alternatively, one can define the thermopower (via an Onsager relation \cite{Ashcroft}) in terms of the heat current $J^Q_x$ produced by a given electrical current $J^e_x$ in situations where the temperature $T$ is uniform. Specifically,
\be
S_{xx}=\frac{1}{T}\frac{J^Q_x}{J^e_x}\label{eqn:1}.
\ee
Throughout this paper we generally describe the thermopower in terms of this latter definition.


In a single-band conductor at low temperature, the Seebeck coefficient is typically of order $(k_B/e) \times k_B T/\vef$, where $k_B$ is Boltzmann's constant, $-e$ is the electron charge, and $\vef$ is the Fermi energy (defined relative to the bottom of the band).  Heuristically, one can think that this small factor $k_B T/\vef$ arises because all electrons in the Fermi sea participate in carrying electric current, while only a small fraction of thermally-excited electrons having energies within $\sim k_B T$ of the Fermi energy carry heat. For this reason, large Seebeck coefficient typically arises only in systems with small Fermi energy, such as doped semiconductors. Unfortunately, low-energy states in semiconductors are prone to localization, which presents a problem for efforts to achieve effective thermoelectrics.\footnote{
Insulators and lightly-doped semiconductors may in fact have relatively large thermopower, proportional to the activation energy divided by $k_B T$ \cite{FRITZSCHE19711813, Chen2013anomalously}. But the exponentially small electrical conductivity in the insulating state typically precludes them from providing efficient power conversion.
}
The recently-discovered three-dimensional Dirac and Weyl semimetals (see, e.g., Refs.\ \cite{armitage_review_2018} and \cite{Hu_review_2019} for reviews) therefore offer significant promise as thermoelectrics (see, e.g., Refs.\ \cite{fiete_thermoelectric_2014, peng2016high, wang2018magnetic, Xiang2019} and Ref.\ \cite{Fu_Felser_review} for a review), since they offer the combination of low Fermi energy, high electrical mobility \cite{shekhar2015extremely, liang2015ultrahigh}, and a gapless electron spectrum that precludes the possibility of localization \cite{skinner_coulomb_2014, syzranov_critical_2015, rodionov_conductivity_2015}.

\begin{figure}
    \centering
    \includegraphics[width=0.8\columnwidth]{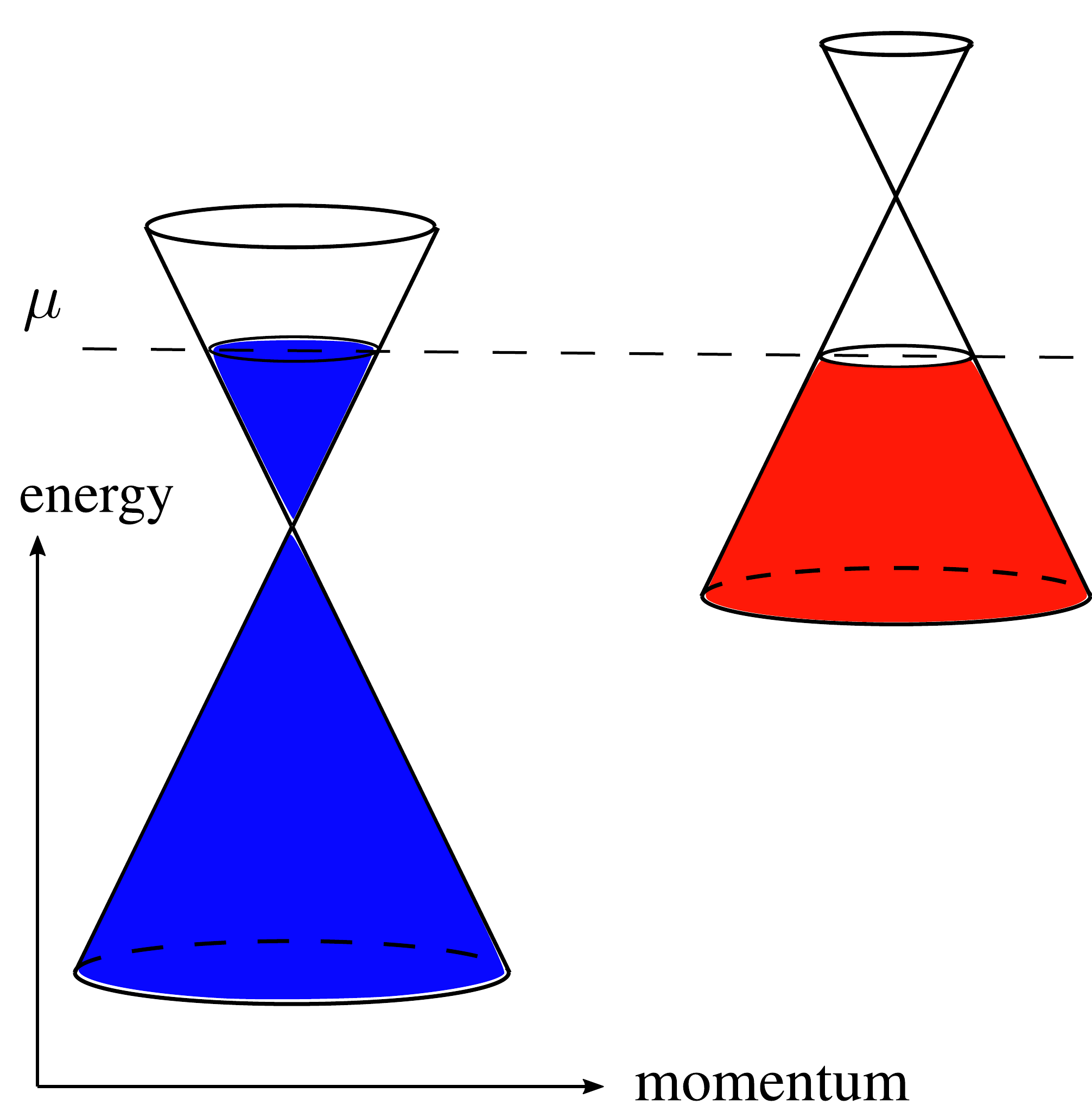}
    \caption{An illustration of the band structure of a compensated Dirac/Weyl semimetal. The band structure has two valleys, which we refer to as the electron valley (blue) and the hole valley (red), with their respective Dirac points offset in energy such that the chemical potential $\mu$ is in the conduction band of the electron valley and the valence band of the hole valley. Shaded areas represent occupied states.}
    \label{fig:1}
\end{figure}

A recent series of papers has shown that the thermopower of a Dirac/Weyl semimetal grows sharply when it is subjected to a sufficiently strong magnetic field that the system reaches the extreme quantum limit (EQL), in which only one Landau level is occupie \cite{Skinnereaat2621, PhysRevB.99.155123, Zhang2020}. Achieving the EQL typically requires a relatively large magnetic field, of order $10\,\textrm{T} \times (n_e\, [10^{17}\, \textrm{cm}^{-3}])^{2/3}$, where $n_e$ is the three-dimensional concentration of electrons. A variety of experiments, however, have demonstrated a large enhancement of thermopower beginning at much smaller magnetic field. For example, a recent experiment in the Weyl semimetal tantalum phosphide (TaP) exhibits an enhancement of $S_{xx}$ by more than two orders of magnitude, starting at a magnetic field of $\approx 0.1$\,T, even though the carrier concentration is of order $10^{19}$\,cm$^{-3}$ \cite{Han2020}. An older experiment in elemental bismuth (a conventional semimetal) demonstrated a similarly huge enhancement of thermopower beginning at low fields \cite{PhysRevB.14.4381}.  In both cases the thermopower reaches values in excess of $1000\, \mu$V/K at cryogenic temperatures, much larger than the naive scale $k_B/e \approx 86 \, \mu$V/K.  
Even more surprisingly, these experiments are in almost-completely-compensated systems, for which pockets of electron- ($e^-$) and hole- ($h^+$) type carriers coexist at the Fermi level (as illustrated in Fig.~\ref{fig:1}) and the corresponding concentrations $n_e$ and $n_h$ of electrons and holes are nearly equal in magnitude. Typically, in such systems the electron and hole contributions nearly cancel in thermopower (unless the two bands have very different mobility \cite{markov_semimetals_2019}), bringing the value of $S_{xx}$ to a small value that is proportional to $(n_e - n_h)/(n_e + n_h)$. 
These experimental results suggest that there is a mechanism for enhancement of the thermopower by magnetic field that is specific to compensated semimetals and does not require the extreme quantum limit. Throughout this paper we neglect the effects of phonon drag, which generally serve to increase the thermopower \cite{ziman_principles_1972, jay-gerin_thermoelectric_1975, Jaoui2020}.

In this paper, we elucidate this mechanism by calculating the Seebeck and Nernst coefficients of compensated semimetals in a magnetic field.
The key idea is that, when the field is large enough that $\omega_c \tau \gg 1$, where $\omega_c$ is the cyclotron frequency and $\tau$ is the transport scattering time, both electrons and holes can contribute additively to the heat current $J^Q_x$ via their motion through the $\vec{E} \times \vec{B}$ drift. On the other hand, the longitudinal conductivity $\sigma_{xx}$ is strongly reduced by the magnetic field, so that the electric current $J^e_x$ is reduced for a given electric field strength. In this way there is a sharp increase in $S_{xx} = J^Q_x/(T J^e_x)$ once the field is large enough that $\omega_c \tau \gg 1$, even though such fields correspond to small Hall angle and are well below the EQL. Indeed, this enhancement mechanism relies on achieving simultaneously large $\omega_c \tau$ and small Hall angle $\theta_H = \arctan\left( \sigma_{xy} / \sigma_{xx} \right)$ (where $\sigma_{xy}$ is the Hall conducitivity). This set of conditions is generally not possible in single band systems. When both conditions are present, however, the thermopower grows as $B^2$, where $B$ is the magnetic field strength.  This $B^2$ enhancement of thermopower is generic for all semimetals with nearly-complete compensation, $|n_e - n_h| \ll n_e + n_h$. In this paper calculate the form of $S_{xx}(B)$ explicitly for both Dirac/Weyl semimetals and for conventional semimetals with parabolic band dispersion.


The Nernst coefficient $S_{xy}$, which describes the off-diagonal thermoelectric response \cite{PhysRevB.97.161404,PhysRevLett.118.136601,PhysRevLett.114.176601}, is also strongly enhanced by the magnetic field. As we show below, in the regime of $\omega_c \tau \gg 1$ and $\sigma_{xy} \ll \sigma_{xx}$ mentioned above, $S_{xy}$ grows linearly with $B$ and achieves a maximum value proportional to $(n_e + n_h)/|n_e - n_h|$.


The remainder of this paper is organized as follows. Section.~\ref{sec:semi-quantitative} gives a semiquantitative derivation of our main result, which is the $B^2$ enhancement of thermopower. Section.~\ref{sec:AD} outlines our calculation method using two complementary approaches: a semiclassical description based on the Boltzmann equation that is valid outside the EQL, and a calculation based on quantum Hall edge states that is valid for all $\omega_c \tau \gg 1$. Sections.~\ref{sec:Weyl} and \ref{sec:semi-conductor} present quantitative results for the Seebeck and Nernst coefficients for compensated Weyl semimetals and compensated semiconductors, respectively. In each of these sections we consider the full range of magnetic field regimes, from arbitrarily small values to deep in the extreme quantum limit.  We conclude in Sec.~\ref{sec:concl} with a summary and discussion.

\section{Semi-quantitative Discussion}
\label{sec:semi-quantitative}

Before giving an exact derivation of the thermopower as a function of $B$, we first present a semiquantitative derivation of the main result in this paper, namely the large enhancement of $S_{xx}$ by magnetic field at $\omega_c \tau \gg 1$. This section includes both a general discussion of different regimes of magnetic field and a conceptual, semiquantitative derivation of $S_{xx}$ in each regime outside the EQL.

\subsection{Regimes of Magnetic Field}
\label{regimes}

In usual conductors with a single band and large Fermi energy $\ve_F$, the two relevant magnetic field regimes for describing transport are $\omega_c \tau \ll 1$ and $\omega_c \tau\gg  1$. Here $\omega_c$ is the cyclotron frequency of electrons at the Fermi energy, which in Dirac/Weyl semimetals increases with Fermi energy. We define the field scale $B_1$ such that at $\omega_c \tau = 1$ at $B = B_1$, which means that $B_1 = m/(e \tau)$ in the usual case of parabolic bands with mass $m$ and $B_1 = \ve_F/(e \tau v_F^2)$ in the case of Dirac/Weyl semimetals with Fermi velocity $v_F$. For simplicity, we assume throughout this paper that $\tau$ is equal for both electron and hole bands. 

In single-band systems $B_1$ is also the magnetic field scale at which $\sigma_{xy}$ becomes comparable to $\sigma_{xx}$, so that $B \gg B_1$ corresponds to large Hall angle $\theta_H$. In nearly-compensated bands, however, the Hall conductivity remains small at $B_1$ due to near-cancellation of electron and hole contributions, so that $\arctan \theta_H(B = B_1) \approx (n_e - n_h)/(n_e + n_h)$.
%
%
The field scale associated with large Hall angle, $\sigma_{xy} \gg \sigma_{xx}$, is therefore significantly larger than $B_1$. We denote the field at which $\sigma_{xy} = \sigma_{xx}$ by $B_H \sim B_1 \times n_e/(\Delta n)$, so that large Hall angle corresponds to $B \gg B_H$. Here $\Delta n = n_e - n_h$ denotes the difference between electron and hole concentrations. Throughout this paper we focus on the case where $\Delta n \ll n_e$, so that $(\Delta n)/n_e$ is a small parameter.

Whenever the magnetic field is low enough that the Landau level spacing $\hbar \omega_c$ is much smaller than the Fermi energy $\ve_F$, one can describe thermoelectric transport in terms of a quasiclassical picture in which the Landau quantization of electron states is relatively unimportant. In the opposite limit of $\hbar \omega_c \gg \ve_F$, however, nearly all electrons reside in the lowest Landau level of transverse motion, and transport must be described in terms of Landau levels. This extreme quantum limit constitutes a different field scale $B_{EQL}$, which equals $2^{1/3}\pi^{4/3}n_e^{2/3}\hbar/(eg^{2/3})$ for Weyl semimetal and $ \pi^{4/3}n_e^{2/3}\hbar/(2^{1/3}eg^{2/3})$ for semiconductor, such that at $B > B_{EQL}$ all electrons reside in the lowest Landau level at zero temperature. We use $2g$ to represent the band degeneracy of each carrier type (including spin degeneracy); for instance, in Dirac/Weyl semimetals, $2g$ represents the number of electron- or hole- type Dirac nodes multiplied by the spin degeneracy. Throughout this paper we assume that $B_{EQL} \gg B_H$, which corresponds to $n_e / (\Delta n) \ll n_e^{1/3} \tau v_F$ for compensated semimetals and $n_e /(\Delta n) \ll \hbar n_e^{2/3}\tau/m$ for parabolic bands, so that Landau quantization effects are relatively unimportant at all but the highest field scales. In Sec.~\ref{sec:concl} we comment briefly on the case where the compensation is so complete that $B_{EQL} \ll B_H$.

\subsection{Mechanism for Large Enhancement of Thermopower}
\label{mechanism-enhancement}

In order to elucidate the mechanism for large field enhancement of the thermopower, we now give a semiquantitative derivation of the Seebeck coefficient $S_{xx}$ in the three semiclassical regimes of magnetic field $B \ll B_1$, $B_1 \ll B \ll B_H$, and $B \gg B_H$.  Discussion of the extreme quantum limit is deferred until the subsequent sections.
As mentioned in Sec.~\ref{sec:Intro}, the Seebeck coefficient can be understood by considering situations in which the temperature is spatially uniform while an electric current flows along the $x$ direction, so that $S_{xx}$ is described by Eq.~(\ref{eqn:2}). 

For the sake of comparison, we begin by considering the usual case of a single band of carriers with concentration $n_e$ (which, for concreteness, we take to be electron-type).
As mentioned in the Introduction, in such cases the thermopower at low magnetic field is of order $S_{xx} \sim (k_B/e) \times k_B T/\ve_F$, where $\ve_F$ is the Fermi energy. 
One can derive this expression in a semiquantitative way by noting that the thermal energy density $U$ at temperature $T$ is of order $U \sim k_B^2 T^2 \nu$, where $\nu$ is the density of states and is typically of order $\nu \sim  n_e/\ve_F$. The heat current density $J^Q_x \sim U v_d$, where $v_d$ is the carrier drift velocity in an applied electric field. Meanwhile, the electric current density $J^e_x \sim -e n_e v_d$.  Combining these expressions gives a thermopower
\be
S_{xx}\sim-\frac{k_B}{e}\frac{k_B T}{\ve_F}. \label{eqn:2}
\ee
A magnetic field $B \ll B_{EQL}$ does not strongly change this result, since the heat current carried by electrons in a single band at low temperature is always proportional to $T^2$ times the electric current, and the field produces only weak modulations of the density of states.

\begin{figure}[htb]
    \centering
    \includegraphics[width=0.9\columnwidth]{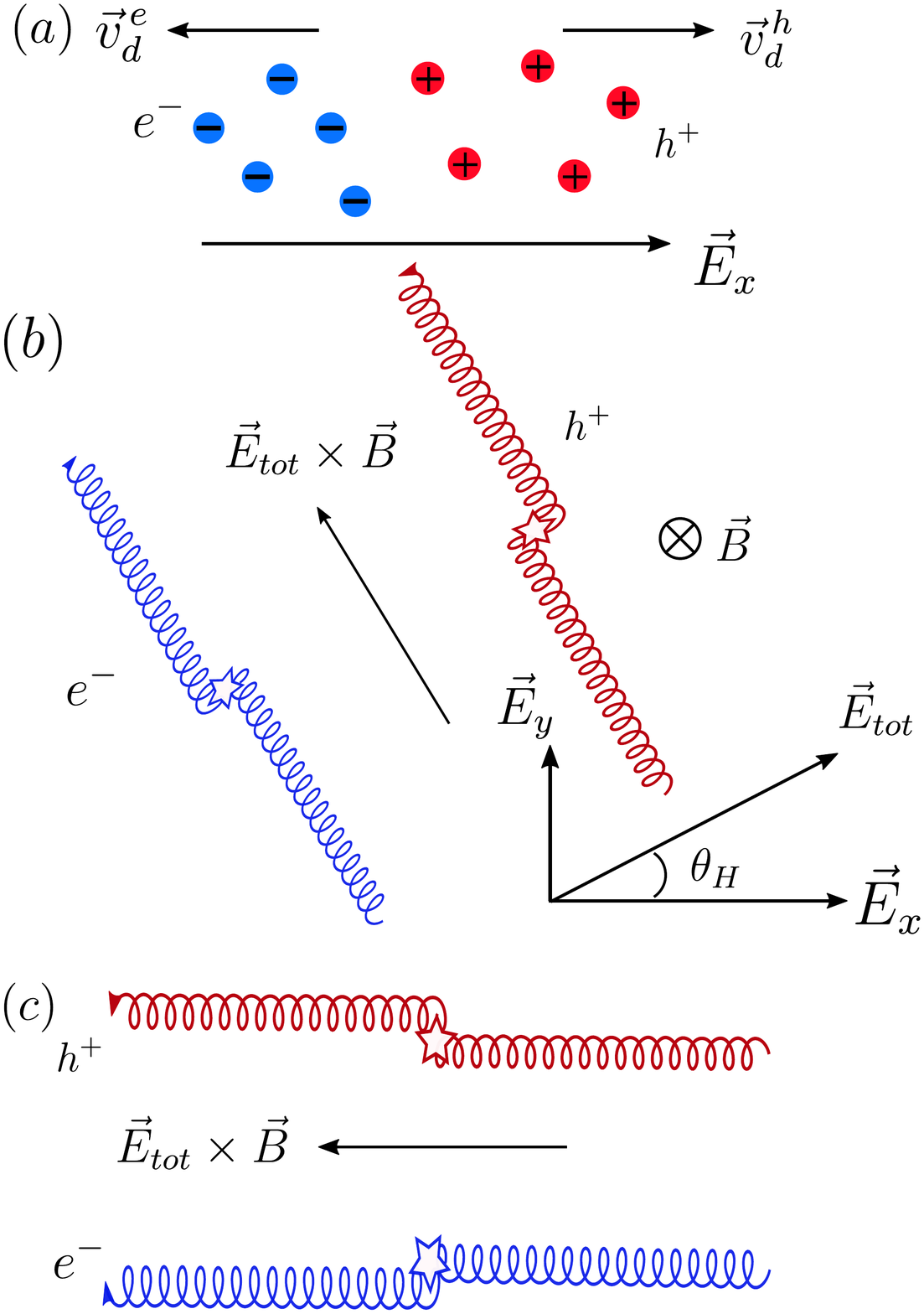}
    \caption{An illustration of the semiclassical motion of electrons and holes in different regimes of magnetic field. (a) At $B \ll B_1$, the magnetic field is negligible and electrons (blue) and holes (red) have opposite drift velocity under the applied electric field, leading to a near-cancellation in the heat current and a small thermopower. (b) When $B_1 \ll B\ll B_H$, the Hall angle $\theta_H$ remains small, but the heat current is dominated by the $x$ component of the $\vec{E} \times \vec{B}$ drift, which allows electron and hole carriers to contribute additively. The star symbols denote impurity scattering events, which limit the electric current. (c) When $B\gg B_H$, the Hall angle $\theta_H$ is nearly $90\degree$ and the $\vec{E}\times \vec{B}$ drift velocity is nearly aligned with the current direction $x$, so that it determines both the heat current and electric current.}
    \label{fig:2}
\end{figure}

In a strongly-compensated system, however, the situation is very different. In the absence of a magnetic field, electrons and holes move in opposite directions under an applied electric field, and therefore they carry heat in opposite directions even as they carry current in the same direction [see Fig.~\ref{fig:2}(a)]. That is, the drift velocity $v_d$ is opposite for electrons and holes, so that the electric current is $J^e_x \sim -e \left(n_e+n_h\right)v_d$. The heat current, on the other hand, is small in magnitude: $J^Q_x \sim k_B^2 T^2  \Delta n/\ve_F$. The resulting Seebeck coefficient
\be
S_{xx}\sim- \frac{k_B}{e} \frac{k_B T}{\ve_F} \frac{\Delta n}{n_e}, \hspace{5mm} B \ll B_1 \label{eqn:3}
\ee
is therefore suppressed by a factor $(\Delta n)/n_e$ relative to the single-band case. This result remains valid for all $B \ll B_1$, for which $\omega_c \tau \ll 1$. 

Now consider the regime of magnetic field {$B_1 \ll B \ll B_H$}. At such fields the Hall conductivity remains small, $\sigma_{xy} \ll \sigma_{xx}$, so that the electric current flows nearly parallel to the applied electric field. The longitudinal conductivity $\sigma_{xx}$, however, declines in this regime as $1/B^2$. In particular, if we define the electrical mobility $\mu_e$, then the Drude formula gives $\sigma_{xx} = e (n_e + n_h) \mu_e / [1 + (\mu_e B)^2] \sim e n_e / (\mu_e B^2)$. (For Dirac/Weyl semimetals, the electric mobility $\mu_e=e\tau \vef/\vf^2$, while for the semiconductor case $\mu_e=e\tau/m$.) Thus, if the electric field has a component $E_x$ in the $x$ direction, then the electric current $J^e_x \sim e n_e E_x / \mu_e B^2$. The corresponding heat current can be found by considering that when the electric current flows along the $x$ direction, there is a $y$ component of electric field $E_y = -\sigma_{xy} E_x / \sigma_{xx} \sim -\mu_e B E_x \Delta n/n_e $. This $y$ component implies that the $\vec{E} \times \vec{B}$ drift velocity has a component in the $x$ direction, which has the same sign for both electrons and holes [see Fig.~\ref{fig:2}(b)]. Multiplying the magnitude of this $x$ component by the internal energy density gives a heat current $J^Q_x \sim - k_B^2T^2 \mu_e E_x \Delta n/\ve_F$. Combining these two results, the Seebeck coefficient is
\be
S_{xx}\sim -\frac{k_B}{e}\frac{k_BT}{\ve_F}\frac{\Delta n}{n_e}\mu_e^2B^2, \hspace{5mm} B_1 \ll B \ll B_H .\label{eqn:4}
\ee
This relation, $S_{xx} \propto [(\Delta n)/n_e] B^2$, is generic for compensated semimetals, regardless of the details of the band dispersion, and is one of the primary results of this paper.

When the magnetic field is further increased to the point that $B\gg B_H$, the Hall angle approaches $90^{\circ}$, and the $\vec{E}\times\vec{B}$ drift velocity becomes nearly aligned with the current direction $x$ [see Fig.~\ref{fig:2}(c)]. In this limit, the flow of current is nearly perpendicular to the electric field direction, and therefore it can be described as a dissipationless process, so that the Seebeck coefficient is described by the simple relation discussed in Refs.\ \cite{jay-gerin_thermoelectric_1974, Skinnereaat2621}: $S_{xx} = (\textrm{total entropy})/(\textrm{net charge})$.  Since, as discussed above, the entropy is of order $k_B^2 T n_e / \ve_F$, the Seebeck coefficient is
\be
S_{xx}\sim -\frac{k_B}{e}\frac{k_BT}{\vef}\frac{n_e}{\Delta n}, \hspace{5mm} B \gg B_H \label{eqn:5}
\ee
Note that this ``saturation'' value of the Seebeck coefficient represents a large enhancement over the value associated with a single-band system [Eq.~(\ref{eqn:2})], by a factor $n_e/(\Delta n)$. 

\begin{figure}
    \centering
    \includegraphics[width=0.9\columnwidth]{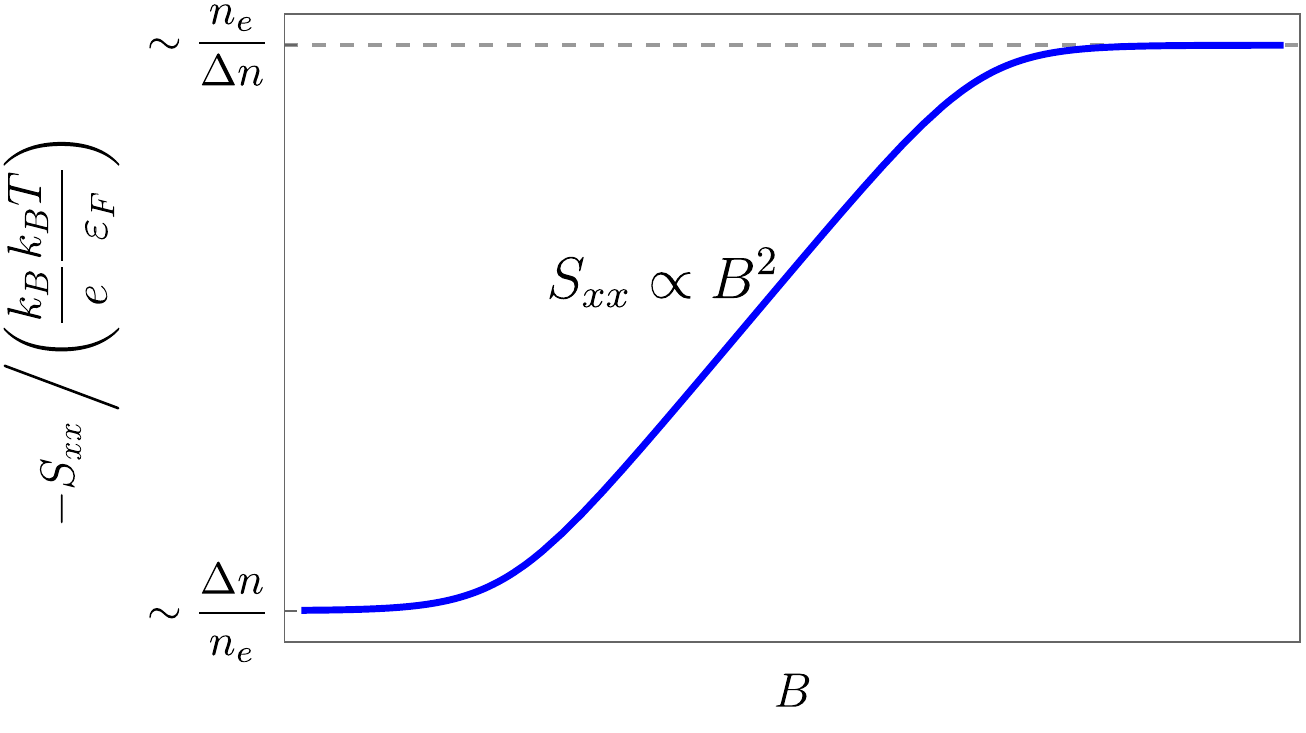}
    \caption{A schematic plot (in double-logarithmic scale) summarizing the three regimes of magnetic field for the Seebeck coefficient in nearly compensated semimetals outside the EQL.}
    \label{fig:3}
\end{figure}

These three regimes are summarized in Fig.~\ref{fig:3}.

It should be emphasized that our arguments in this section have focused on the semiclassical regime $B \ll B_{\text{EQL}}$, in which many Landau levels are occupied. When $B \gg B_{\text{EQL}}$ this semiclassical description fails, and it should be replaced by a calculation in terms of quantum Hall-type edge states; we discuss such a calculation in Sec.~\ref{sec:dissipationless}. 

One can also ask about the Nernst coefficient $S_{xy}$, which is the off-diagonal component of the thermoelectric tensor [defined as $S_{xy} = (\Delta V)_y / (\Delta T)_x$, or $S_{xy} = J^Q_y / (T J^e_x)$]. Similar semiquantitative arguments give 
\begin{align}
S_{xy}  \sim &\frac{k_B}{e} \frac{k_B T \mu_e B}{\vef},  & B \ll B_H, \\
S_{xy}  \sim &\frac{k_B}{e} \frac{k_B T n_e^2}{(\Delta n)^2 \mu_e B\vef},  & B \gg B_H,
\end{align}
so that $S_{xy}$ achieves a large maximum value proportional to $n_e / (\Delta n)$ at $B \sim B_H$.

\section{Analytical Description}
\label{sec:AD}

\subsection{Semi-classical Theory}
\label{sec:st}

At low temperature, $k_B T \ll \ve_F$, the thermoelectric tensor $\hat{S}$ can be calculated by the Mott formula~\cite{Ashcroft},
\be
\hat{S}=-\frac{\pi^2}{3}\frac{k_B}{e}k_BT\hat{\sigma}^{-1}\left.\frac{d\hat{\sigma}}{d\ve}\right|_{\ve_F}.\label{eqn:6}
\ee
Thus, the Seebeck and Nernst coefficients are completely defined by the relationship between the conductivity tensor $\hat{\sigma}$ and the energy $\ve$ at zero temperature. 
In the semi-classical regime $B \ll B_{EQL}$, this relationship can be obtained from the Boltzmann equation, which we briefly recapitulate here. 

For the case with both electrons and holes, the total conductivity is \cite{Ashcroft}
\be
\hat{\sigma}=\hat{\sigma}^e+\hat{\sigma}^h,\label{eqn:7}
\ee
where $\hat{\sigma}^e$ and $\hat{\sigma}^h$ are the electron and hole conductivity tensors, respectively, given by
\be 
\hat{\sigma}^{e,h}\left(\ve\right)=e^2\tau(\ve)\int \frac{d\mathbf{k}}{4\pi^3}\delta\left(\ve-\ve(\mathbf{k})\right)\mathbf{v}^{e,h}\left(\mathbf{k}\right)\Bar{\mathbf{v}}^{e,h}\left(\mathbf{k}\right),\label{eqn:8}
\ee
with
\be
\Bar{\mathbf{v}}\left(\mathbf{k}\right)=\int_{-\infty}^{0}\frac{dt}{\tau(\ve)}\,e^{t/\tau(\ve)}\,\mathbf{v}\left(\mathbf{k}\left(t\right)\right) \label{eqn:9}
\ee
and
$\mathbf{v}(\mathbf{k})$ denoting the group velocity $\mathbf{v}(\mathbf{k})=\nabla_{\mathbf{k}}\ve(\mathbf{k})$. The time evolution of the momentum is given by the semiclassical equation of motion
\be
\hbar\dot{\mathbf{k}}=\mp e\, \mathbf{v}\left(\mathbf{k}\right)\times \mathbf{B}.\label{eqn:10}
\ee
The dependence of the scattering time $\tau$ on the quasiparticle energy $\ve$ depends in general on the scattering mechanism, and can have a variety of different functional forms. For the sake of our discussion in this paper, we assume that $\tau$ is a constant and independent of $\ve$. If one includes an energy dependence for $\tau$, then certain numerical prefactors are modified in formulas containing $\tau$, but our primary results are unchanged.

With these assumptions, one can derive the conductivity tensors as
\be
\hat{\sigma}^e\left(\ve\right)=\frac{1}{3}\frac{e^2\nu_e\left(\ve\right)v_e^2\left(\ve\right)\tau}{1+\omega_c^2\tau^2}
\begin{pmatrix}
1&\omega_c\tau\\-\omega_c\tau&1
\end{pmatrix},\label{eqn:11}
\ee
\be
\hat{\sigma}^h\left(\ve\right)=\frac{1}{3}\frac{e^2\nu_h\left(\ve\right)v_h^2\left(\ve\right)\tau}{1+\omega_c^2\tau^2}
\begin{pmatrix}
1&-\omega_c\tau\\\omega_c\tau&1
\end{pmatrix}.\label{eqn:12}
\ee
Together with the Mott formula [Eq.~(\ref{eqn:6})], Eqs.\ (\ref{eqn:11}) and (\ref{eqn:12}) completely define the Seebeck and Nernst coefficients at temperatures $T \ll \ve_F/k_B$.

\subsection{Dissipationless Limit}
\label{sec:dissipationless}

When the magnetic field is large enough that $\sigma_{xy} \gg \sigma_{xx}$, one can describe the thermopower in the dissipationless limit, in which $\tau$ is effectively set to infinity.  In this description, all electrical and thermal current is carried by quantum Hall-type edge states \cite{Girvin_1982,PhysRevB.25.2185}. For nearly compensated systems, the dissipationless limit requires $B\gg B_H$, but it encompasses both the limit where many Landau levels are occupied and the extreme quantum limit $B \gg B_{\text{EQL}}$.

The flow of electrical and thermal current is described by the coupled transport equations~\cite{Ashcroft}, 
\be
\begin{pmatrix}J^e\\J^Q
\end{pmatrix}=\begin{pmatrix}\hat{\sigma}&-\hat{\alpha}\\\hat{\alpha}T&-\hat{\kappa}
\end{pmatrix}\begin{pmatrix}
E\\\nabla T
\end{pmatrix}.\label{eqn:13}
\ee
where $\hat{\kappa}$ is the thermal conductivity tensor and the tensor $\hat{\alpha}$ is 
related to the thermoelectric tensor $\hat{S}$ by $\hat{S}=\hat{\sigma}^{-1} \hat{\alpha}$. (Here we have written the transport coefficients in terms of intensive quantities, rather than in terms of extensive differences in voltage and temperature, as in the Introduction.)
The Seebeck coefficient is its diagonal term, which one can write as
\be
S_{xx}=\frac{\sigma_{xx}\alpha_{xx}+\sigma_{xy}\alpha_{xy}}{\sigma_{xx}^2+\sigma_{xy}^2}.\label{eqn:14}
\ee
In the dissipationless limit ($\tau \rightarrow \infty$), one has $\sigma_{xy}\gg\sigma_{xx}$ and $\alpha_{xy}\sigma_{xy} \gg \alpha_{xx}\sigma_{xx}$, so that the Seebeck coefficient is given simply by
\be
S_{xx}=\frac{\alpha_{xy}}{\sigma_{xx}}.\label{eqn:15}
\ee
Both $\sigma_{xy}$ and $\alpha_{xy}$ are well defined in the limit $\tau \rightarrow \infty$.

In compensated systems, $\sigma_{xy}$ and $\alpha_{xy}$ are given by the sum of contributions from both electron valley and hole valleys. The contribution from each valley can be calculated independently. For concreteness, in the remainder of this section we concentrate on the electron valley, briefly repeating the derivation for the single-band case as presented in Ref.\ \cite{PhysRevB.99.155123}. The hole valley is completely analogous. 

We start by considering a Hall brick with length $L_x$, $L_y$, and $L_z$, respectively. The magnetic field is assumed to be along the $z$ direction and the electric field is assumed to be along the $x$ direction. The Landau gauge is chosen, with vector potential $\vec{A} = \left(0,Bx,0\right)$. It is safe to assume that each Landau level is constant in energy in the bulk along the $x$ direction and increases sharply at the edge of the Hall brick. The contribution of the electron valley to the electric current along the $y$ direction is
\be
I_y=-\frac{e}{L_y}\sum_\textrm{states}v_y n_F\left(\ve-\mu\right),\label{eqn:16}
\ee
where $n_F$ is the Fermi-Dirac distribution and $\mu$ is the chemical potential. Given that the electric field is small, the Fermi-Dirac distribution can be expanded to first order in the potential difference $\Delta V_x$ along the $x$ direction. 
The $n$th Landau level $\epsilon_n(k_y,k_z)$ in the bulk is almost flat (independent of $k_y$) and gives little contribution to the current. Only the contribution of edge states $k_y=\pm L_x/2l_B^2$ needs to be included, where $l_B = \sqrt{\hbar/(eB)}$ is the magnetic length.  The corresponding electric current
\begin{align}
I_y=&\frac{e^2L_z\Delta V_x}{2\pi\hbar}\int_{-\infty}^{\infty}\frac{dk_z}{2\pi}\times\left[\sum_{\ve_n>0} N_n n_F\left[\ve_n(k_z)-\mu\right]\right.\nonumber\\
&\left.-\sum_{\ve_n<0}N_n\left(1-n_F\left[\ve_n(k_z)-\mu \right]\right)\right],\label{eqn:17}
\end{align}
where $N_n$ is the degeneracy of the $n$th Landau level at a given momentum $k_z$. For a given $n_e$, the chemical potential $\mu$ is fixed by the condition
\begin{align}
&\int_{0}^{\infty}d\ve\frac{eB}{2\pi\hbar}\sum_{k_z,n}N_n\,\delta(\ve-\ve_n(k_z))n_F\left[\ve-\mu\right]\nonumber\\
&\quad+\int_{\infty}^0d\ve\frac{eB}{2\pi\hbar}\sum_{k_z,n}N_n\delta\left(\ve-\ve_n(k_z)\right)\left(1-n_F\left[\ve-\mu\right]\right)=n_e.\label{eqn:18}
\end{align}
 The resulting Hall conductivity of the electron valley
\be
\sigma^e_{xy}= - \frac{en_e}{B}.\label{eqn:19}
\ee

The heat current in the electron valley is obtained in a similar way,
\begin{align}
I^Q_y=&-\frac{e}{\hbar}\frac{\Delta V_xl_B^2}{L_xL_y}\sum_{k_z,k_y,n}N_nk_y\,\left[\ve_n(ky,k_z)-\mu\right]\nonumber\\
&\times\frac{\partial\ve_n\left(k_y,k_z\right)}{\partial k_y}\frac{\partial}{\partial\ve}n_F\left[\ve_n\left(k_y,k_z\right)-\mu\right].\label{eqn:20}
\end{align}
The thermoelectric Hall conductivity is defined by
\begin{eqnarray}
\alpha^e_{xy} & = & \frac{I_y^Q}{T\Delta V_x L_y} \label{eqn:21} \\
& = & \frac{e}{2\pi \hbar L_z}\sum_{n,k_z}N_n s\left(\frac{\ve_n(k_z)-\mu}{k_BT}\right).  \nonumber
\end{eqnarray}
The function $s(x)$ represents the average entropy per electron for a given quantum state,
\be
s\left(x\right)=-k_B\left[n_F(x)\ln n_F(x)+\left(1-n_F(x)\right)\ln\left(1-n_F(x)\right)\right].\label{eqn:22}
\ee
The reader is referred to Ref.\ \cite{PhysRevB.99.155123} for a more detailed presentation.

As mentioned before, the discussion containing Eqs.~(\ref{eqn:16})--(\ref{eqn:22}) focused on the electron valley. But this calculation can be repeated to get the corresponding contributions $\sigma^h_{xy}$ and $\alpha^h_{xy}$ for the hole valley. The total Seebeck coefficient is given by
\be
S_{xx}=\frac{\alpha_{xy}^e+\alpha_{xy}^h}{\sigma_{xy}^e+\sigma_{xy}^h}.\label{eqn:23}
\ee

\section{Compensated Dirac/Weyl Semimetals}
\label{sec:Weyl}

We now present results for the Seebeck and Nernst coefficients for compensated Dirac/Weyl semimetals, using the two complementary calculations outlined in the previous subsections.
As mentioned above, in Dirac/Weyl semimetals the cyclotron frequency $\omega_c = e Bv_F^2/\ve$ depends on the energy $\ve$ relative to the Dirac point. Since the Mott formula is defined in terms of the zero-temperature conductivity, we need only consider the cyclotron frequency at the Fermi energy $\ve = \ve_F$. In the remainder of this section we use $\omega_c$ to denote this value. 

Plugging Eqs.\ (\ref{eqn:11}) and (\ref{eqn:12}) into the Mott formula, Eq.~(\ref{eqn:6}), gives for the Seebeck coefficient
\begin{alignat}{3}
S_{xx}\approx &-\frac{\pi^{4/3}g^{1/3}}{3^{7/3}}\frac{k_B}{e}\frac{k_BT\Delta n}{\hbar \vf n_e^{4/3}},\quad&& B\ll B_1,\label{eqn:24}\\
S_{xx}\approx&-\frac{g}{6}\frac{k_B}{e}\frac{k_BT\Delta ne^2B^2\tau^2v_F}{n_e^2\hbar^3},\quad&& B_1\ll B\ll B_H,\label{eqn:25}\\
S_{xx}\approx&-\frac{2\pi^{4/3}g^{1/3}}{3^{1/3}}\frac{k_B}{e}\frac{k_BTn^{2/3}_e}{\hbar\vf \Delta n},\quad&& B_H\ll B\ll B_{\text{EQL}},\label{eqn:26}
\end{alignat}
where the band degeneracy $g$ is equal to the number of Dirac nodes. Notice that these three regimes are equivalent to the ones discussed Sec.~\ref{sec:semi-quantitative}.

\begin{figure}
    \centering
    \includegraphics[width=0.9\columnwidth]{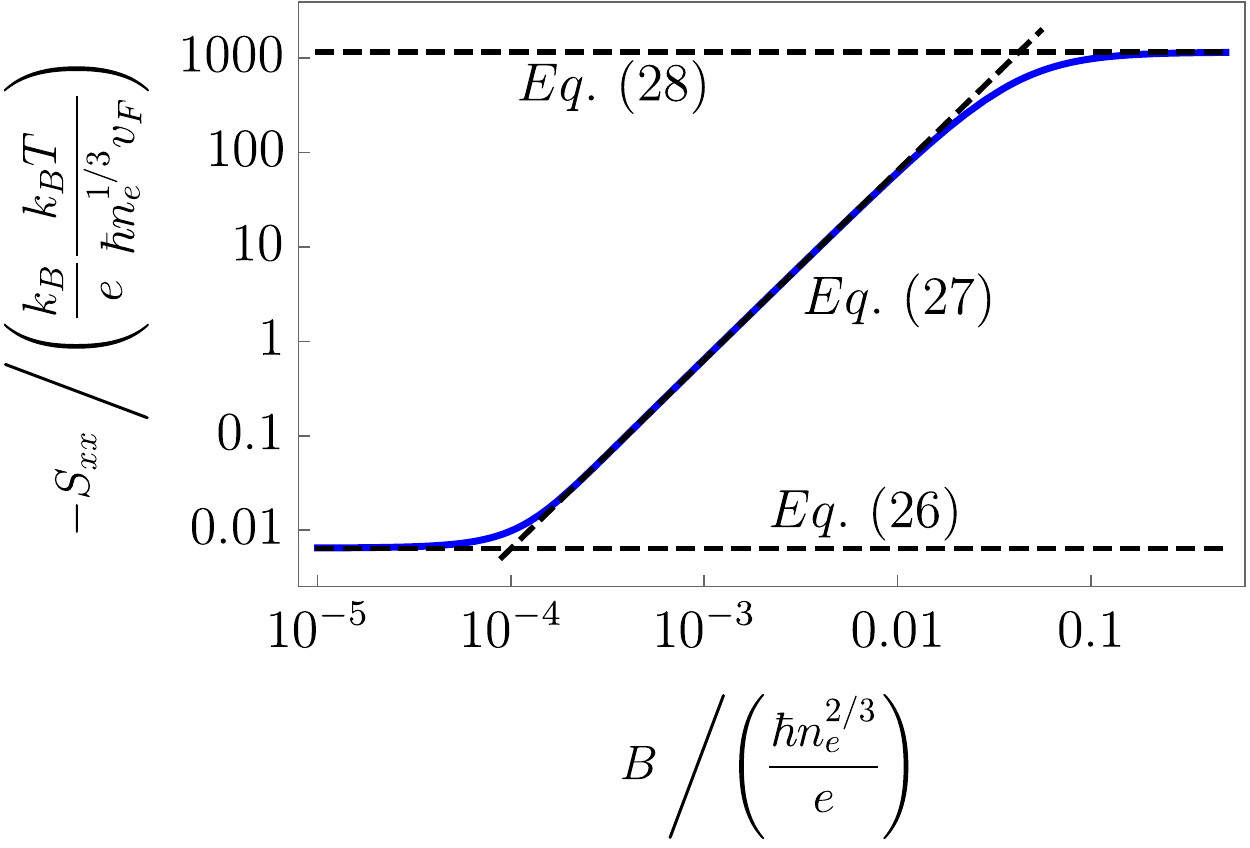}
    \caption{The Seebeck coefficient in a nearly compensated Dirac/Weyl semimetal as a function of $B$, plotted in dimensionless units and in double-logarithmic scale. The magnetic field $B \ll B_{\text{EQL}}$ everywhere in this plot. The scattering time and degree of compensation are such that $\tau=8000 v_F^{-1}n_e^{-1/3}$ and $\Delta n =0.01 n_e$. The band degeneracy $g=6$.  Each regime of magnetic field is labeled by the corresponding asymptotic equation that describes it (dashed lines). Compare the three regimes derived semiquantitatively in Sec.~\ref{sec:semi-quantitative}.}
    \label{fig:4}
\end{figure}

In Fig.~\ref{fig:4} we plot the Seebeck coefficient obtained using the Mott formula, together with the asymptotic expressions of Eqs.\ (\ref{eqn:24})--(\ref{eqn:26}). We use dimensionless units for the values of $B$ and $S_{xx}$ such that the curve $S_{xx}(B)$ is parameterized by only three dimensionless constants: the degeneracy $g$, the scattering time $\tau$ in units of $1/(v_F n_e^{1/3})$ (which is of the order of the Fermi time), and the relative compensation $\Delta n / n_e$.

When the magnetic field is sufficiently large that $\sigma_{xy} \gg \sigma_{xx}$, which encompasses both the regime of Eq.\ (\ref{eqn:26}) and  $B \gg B_{\text{EQL}}$, the current flow becomes nearly dissipationless, and one can calculate the thermoelectric tensor using the picture of dissipationless edge states (Sec.~\ref{sec:dissipationless}). The Landau levels in the bulk of a Dirac/Weyl semimetal are given by \cite{Jeon2014}
\be
\ve_n\left(k_z\right)=\vf\,\sign(n)\sqrt{\hbar^2k_z^2+2e\hbar B\left|n\right|},\label{eqn:27}
\ee
where $n$ is the Landau level index and $\hbar k_z$ is the momentum in the field direction.  Inserting this spectrum into Eqs.~(\ref{eqn:21}) and (\ref{eqn:23}) gives for the Seebeck coefficient
\begin{alignat}{3}
S_{xx}\approx &-\frac{2\pi^{4/3}g^{1/3}}{3^{1/3}} \frac{k_B}{e} \frac{k_B Tn_e^{2/3}}{\Delta n \vf\hbar},\quad && B_H\ll B\ll B_{\text{EQL}},\label{eqn:28}\\
S_{xx}\approx&-\frac{g}{3}\frac{k_B}{e}\frac{k_BT e B}{\hbar^2\vf\Delta n},\quad &&B\gg B_{\text{EQL}}.\label{eqn:29}
\end{alignat}
\par
Notice that Eq.~(\ref{eqn:28}) agrees exactly with the semiclassical result in Eq.~(\ref{eqn:26}). Equation~(\ref{eqn:29}) indicates a linear-in-$B$ enhancement of thermopower in the EQL, as first derived in Ref.~\cite{Skinnereaat2621}. Note, however, that the value of $S_{xx}$ in the EQL is enhanced relative to the single-band case by a large factor $\sim n_e/(\Delta n)$.



\begin{figure}[htb]
    \centering
    \includegraphics[width=0.9\columnwidth]{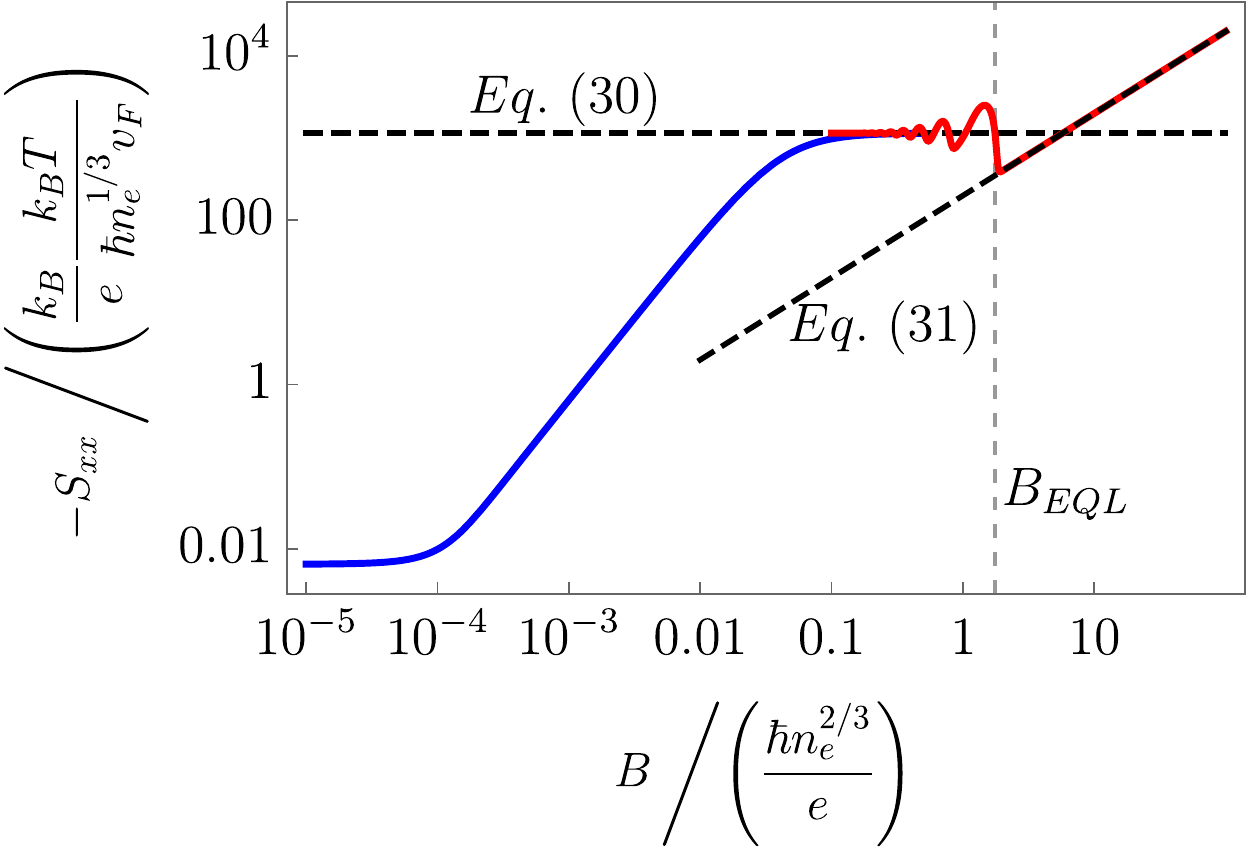}
    \caption{The Seebeck coefficient of a nearly-completely-compensated Dirac/Weyl semimetal as a function of magnetic field, showing both the semiclassical (blue) and dissipationless limit (red) calculations. The material parameters are taken to be the same as in Fig.~\ref{fig:4}. The two asymptotic results applicable to the dissipationless limit are shown as dashed lines.}
    \label{fig:5}
\end{figure}

In Fig.~\ref{fig:5}, the red curve shows the Seebeck coefficient calculated via quantum Hall edge states, and the blue curve shows the semiclassical calculation. The two results match in the regime of magnetic field $B_H \ll B \ll B_{EQL}$.  


The Mott formula also allows us to calculate the Nernst coefficient $S_{xy}$ in the semiclassical limit. This calculation gives:
\begin{alignat}{2}
S_{xy}\approx & \ \frac{4\pi^{2/3}g^{2/3}}{3^{5/3}}\frac{k_B}{e}\frac{k_BTeB\tau }{\hbar^2n_e^{2/3}},\quad&&B\ll B_1,\label{eqn:30}\\
S_{xy}\approx & \ \frac{\pi^{2/3}g^{2/3}}{3^{2/3}}\frac{k_B}{e}\frac{k_BTeB\tau}{\hbar^2n_e^{2/3}},\quad&& B_1\ll B\ll B_{H},\label{eqn:31}\\
S_{xy}\approx & \ 4\pi^2\frac{k_B}{e}\frac{k_BT n_e^2}{(\Delta n)^2 eB v_F^2\tau},	\quad&& B\gg B_{H}.\label{eqn:32}
\end{alignat}

\begin{figure}
    \centering
    \includegraphics[width=0.9\columnwidth]{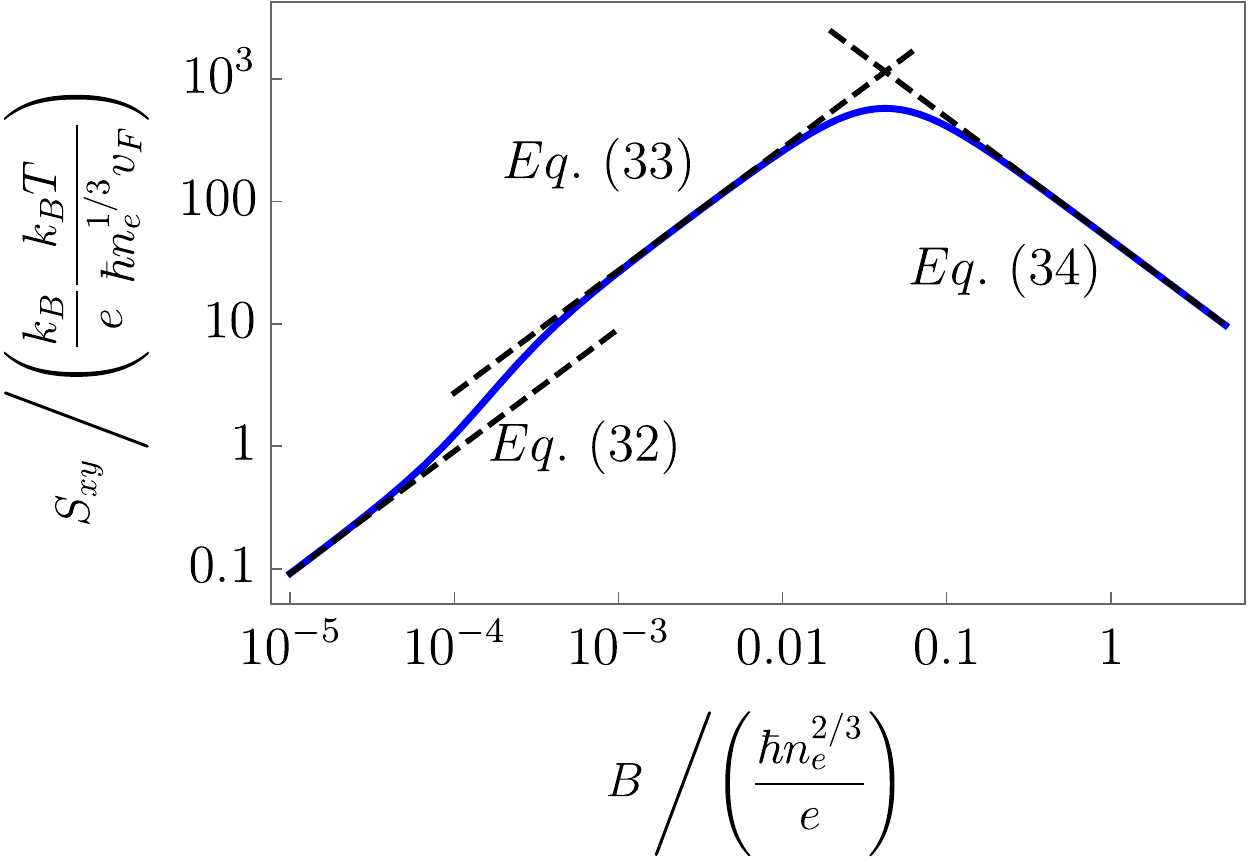}
    \caption{The Nernst coefficient $S_{xy}$ of a nearly-completely-compensated Dirac/Weyl semimetal as a function of magnetic field. The material parameters are taken to be the same as in Figs.\ \ref{fig:4} and \ref{fig:5}. The Nernst coefficient achieves a peak value of order {$S_{xy} \sim (k_B/e)(k_BT/\ve_F)\times n_e/(\Delta n)$} when $B \sim B_H$.}
    \label{fig:6}
\end{figure}

These formulas imply that $S_{xy}$ grows linearly with magnetic field at $B \ll B_H$, achieving a maximum value $S_{xy} \sim (k_B/e) \times (k_B T/\ve_F) \times n_e/(\Delta n)$ at $B \sim B_{H}$, and then declines again as $1/B$ when $B \gg B_H$. These behaviors are shown in Fig.~\ref{fig:6}.
At magnetic fields $B \gg B_{EQL}$, the value of the Nernst coefficient depends on the details of the relevant scattering processes, and it is not well-defined in the dissipationless limit. We therefore leave analysis of $S_{xy}$ in the EQL to a later work.

\begin{figure}
    \centering
    \includegraphics[width=0.9\columnwidth]{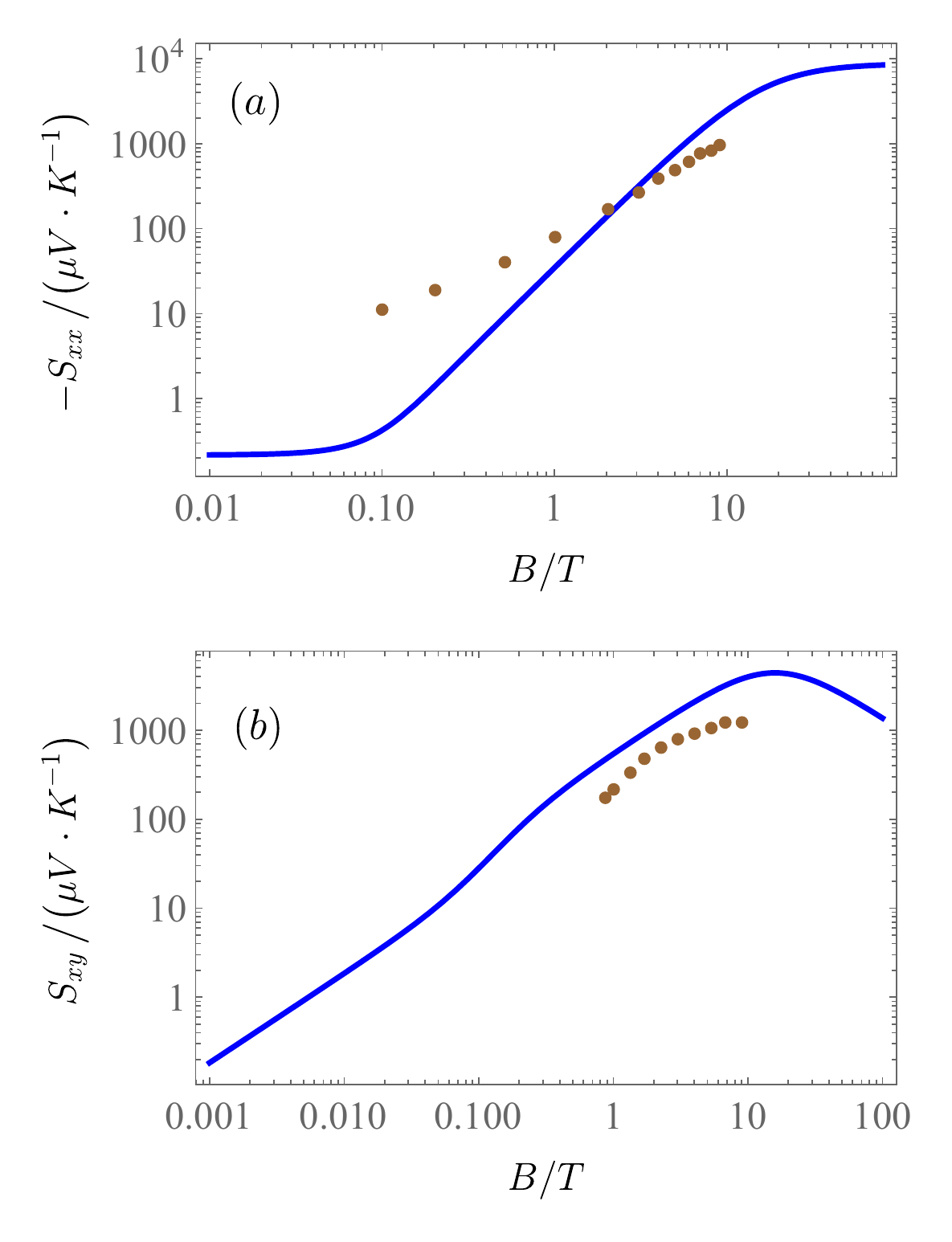}
    \caption{A comparison between the Seebeck (a) and Nernst (b) coefficients calculated in this work (blue lines) and the values reported in Ref.\ \cite{Han2020} for the compensated Weyl semimetal TaP (brown points). The material parameters used for the calculation are taken from Refs.~\cite{Han2020} and \cite{PhysRevB.92.235104}, and there are no fitting parameters. The data corresponds to a temperature $T=50\text{K}$. The lack of quantitative agreement, particularly at small magnetic field, may arise in part from nonlinearity of the dispersion relation in TaP.}
    \label{fig:7}
\end{figure}

The results in this section can be compared to a recent experimental work \cite{Han2020}, which demonstrates an enormous enhancement of Seebeck and Nernst coefficients as a function of magnetic field in the compensated Weyl semimetal TaP. TaP has 12 pairs of Weyl nodes, 4 of which are at energies below the chemical potential (electron type) and 8 of which are above (hole type). In order to make a rough, but quantitative, comparison to the experiment, we use the parameters reported in Ref.\ \cite{Han2020} for the electron density, $n_e=2.4\times 10^{19}$\, cm$^{-3}$, hole density, $n_h=2.35\times 10^{19}$\,cm$^{-3}$, and transport scattering time $\tau=9.76\times 10^{-12}$\,s. Since our results depend only weakly on the band degeneracy, we use $g = 6$ for both electron- and hole-type pockets. For the Fermi velocity, we use geometric of the three orthogonal Fermi velocities reported in Ref.~\cite{PhysRevB.92.235104}.  The resulting calculation is compared to experimental data in Fig.~\ref{fig:7}, using data corresponding to $T = 50$\,K.  There are no fitting parameters in the calculation.

As can be seen in Fig.~\ref{fig:7}, our calculation captures both the order of magnitude of the experimental result and the qualitative trend of strongly increasing $S_{xx}$ and $S_{xy}$. However, the agreement is not very strong, particularly at small magnetic field. This deviation may arise in part from the nonlinearity of the dispersion relation $\ve(\vec{k})$ in TaP, for which the dispersion is only ``Weyl-like'' at energies very close to the Weyl points.  A more accurate calculation that is specific to TaP is beyond the scope of this paper.

\section{Compensated Semiconductors}
\label{sec:semi-conductor}

So far, we have focused primarily on compensated Dirac/Weyl semimetals, but the general mechanism for field enhancement of thermopower outlined in Sec.~\ref{sec:semi-quantitative} is generic to any compensated system. In order to demonstrate this generality, in this section we consider the case of a compensated semiconductor. As an example, we examine the simple situation in which two parabolic bands with identical effective mass $m$, one electron type and one hole type, intersect the chemical potential with nearly identical Fermi energy. In this case the cyclotron frequency $\omega_c=eB/m$ is a constant that does not depend on energy.

\begin{figure}[htb]
    \centering
    \includegraphics[width=0.9\columnwidth]{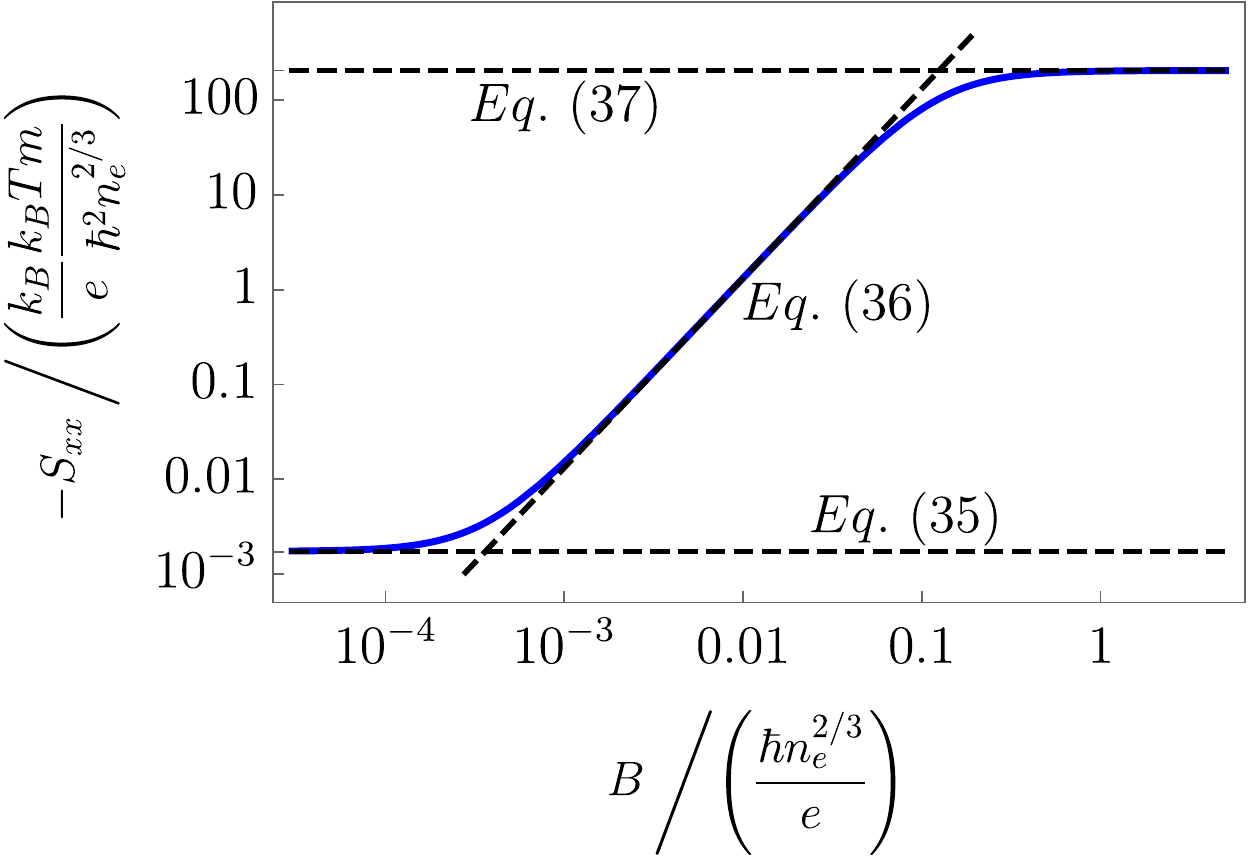}
    \caption{The Seebeck coefficient $S_{xx}$ for a nearly-completely-compensated semiconductor as a function of magnetic field, plotted in double-logarithmic scale. The material parameters used for this calculation are $g = 1$, $\tau=1600 m/(\hbar n_e^{2/3})$, and $\Delta n=0.01 n_e$. The range of field values in this plot correspond to $B < B_{\text{EQL}}$, at which the semiclassical description is valid. Compare the three regimes described in Sec.~\ref{sec:semi-quantitative}.}
    \label{fig:8}
\end{figure}

Using the Mott formula, ~(\ref{eqn:6}), we calculate the Seebeck coefficient in each of the three semiclassical field regimes as
\begin{alignat}{3}
S_{xx}\approx& -\frac{\pi^{2/3}g^{2/3}}{2\times 3^{5/3}}\frac{k_B}{e}\frac{k_BTm\Delta n}{\hbar^2 n_e^{5/3}},&&\quad B\ll B_1,\label{eqn:33}\\
S_{xx}\approx&-\frac{\pi^{2/3}g^{2/3}}{2\times 3^{2/3}}\frac{k_B}{e}\frac{k_BT\Delta ne^2\tau^2B^2}{m\hbar^2n_e^{5/3}},&&\quad B_1\ll B\ll B_H,\label{eqn:34}\\
S_{xx}\approx&-\frac{2\pi^{2/3}g^{2/3}}{3^{2/3}}\frac{k_B}{e}\frac{k_BTmn_e^{1/3}}{\hbar^2\Delta n},&&\quad B_H\ll B\ll B_{\text{EQL}}.\label{eqn:35}
\end{alignat}
Here, $g$ is the degeneracy of each band (including spin). These results are equivalent to the three regimes outlined in Sec.\ \ref{sec:semi-quantitative}. A full semiclassical calculation is presented in Fig.~\ref{fig:8}, along with the relevant asymptotic expressions. The units of our calculation are such that the curve $S_{xx}(B)$ depends only on three dimensionless material parameters: the degeneracy $g$, the transport scattering time $\tau$ in units of $m/(\hbar n_e^{2/3})$, and the degree of compensation $(\Delta n)/n_e$. These three semiclassical regimes are plotted in Fig.~\ref{fig:8}.

In the dissipationless limit $B \gg B_H$, the Seebeck coefficient can be described in terms of quantum Hall edge states, as outlined in Sec.\ \ref{sec:dissipationless}.  The Landau levels for Shr\"{o}dinger particles
satisfy
\be
\ve_n\left(k_z\right)=\frac{\hbar^2k_z^2}{2m}+\hbar\omega_c\left(n+\frac{1}{2}\right), \label{eqn:36}
\ee
where $n$ is the Landau level index and $\hbar k_z$ is the momentum in the magnetic field direction. Using Eqs.~(\ref{eqn:15}), (\ref{eqn:19}), and (\ref{eqn:21}), one can derive the  Seebeck coefficient as
\begin{alignat}{3}
S_{xx}\approx& -\frac{2\pi^{2/3}g^{2/3}}{3^{2/3}}\frac{k_B}{e}\frac{k_BTm n_e^{1/3}}{\hbar^2\Delta n},  &&B_H\ll B\ll B_{EQL},\label{eqn:37}\\
S_{xx}\approx&-\frac{2g^2}{3\pi^2}\frac{k_B}{e}\frac{e^2k_BTmB^2}{\hbar^2n_e\Delta n},  &&B_{EQL}\ll B\ll B_T,\label{eqn:38}\\
S_{xx}\approx &-2\frac{k_B}{e}\frac{n_e }{\Delta n}\ln\left ( B/B_T\right), &&B\gg B_T.\label{eqn:39}
\end{alignat}
Here we have defined a new field scale 
$B_T = \sqrt{2} \pi \hbar^2 n_e / (g e \sqrt{m k_B T})$,
such that at $B \gg B_T$ the Fermi energy (relative to the bottom of the conduction band) becomes smaller than $k_B T$. 

Equation (\ref{eqn:37}) is identical to the semiclassical result in Eq.~(\ref{eqn:35}), and corresponds to the limit where many Landau levels are occupied.  Once the EQL is reached, and only a single Landau level is occupied, the Fermi energy (relative to the minimum energy $\hbar \omega_c/2$ of the conduction band) begins to fall with increased field as $1/B^2$, which is a consequence of the strongly enhanced density of states in the lowest Landau level \cite{PhysRevB.99.155123,Bhattacharya2016}.  The $B^2$ enhancement of the thermopower at $B_{EQL} \ll B \ll B_T$ reflects this falling Fermi energy, and the associated rise of the fraction of thermally-excited electrons.   However, when the Fermi energy falls so far that it becomes much smaller than $k_B T$, the chemical potential falls into the band gap and the electron energies are well-described by a classical Boltzmann distribution.  The electron entropy, which determines the thermopower in the dissipationless limit, is therefore given by an analog of the Sackur-Tetrode equation for the entropy of an ideal gas, leading to the logarithmic dependence in Eq.~(\ref{eqn:39}) \cite{PhysRevB.99.155123}. This logarithmic regime $B \gg B_T$ does not exist in the Dirac/Weyl case, because there is no band gap and therefore no regime in which the electrons obey classical, Boltzmann statistics.

\begin{figure}
    \centering
    \includegraphics[width=0.9 \columnwidth]{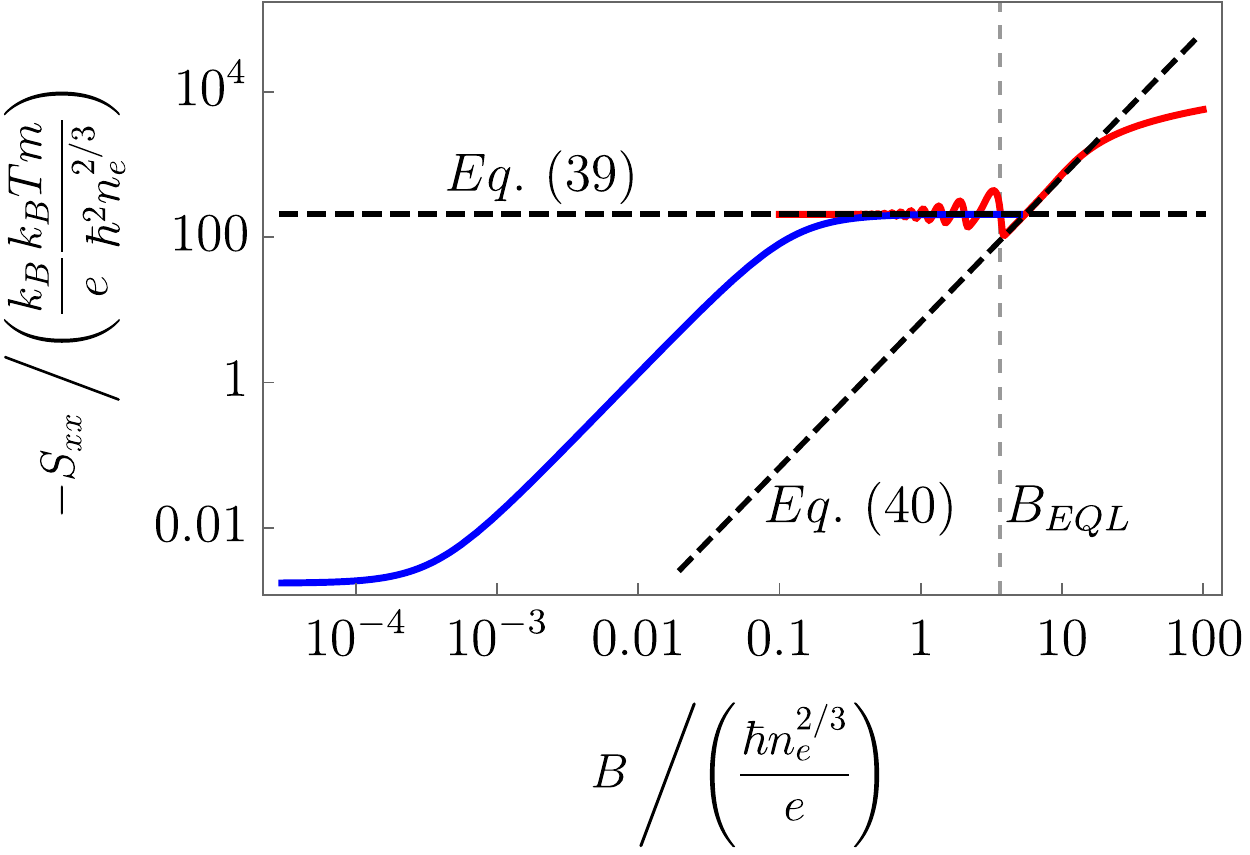}
    \caption{The Seebeck coefficient $S_{xx}$ for a nearly-completely-compensated semiconductor as a function of magnetic field $B$, including all regimes of $B$. The material parameters are the same as in Fig.~\ref{fig:8}. The temperature is chosen such that $T = 0.1 \hbar^2n_e^{2/3}/(mk_B)$; in the units of our plot this choice affects only the largest field regime $B \gg B_T$. The blue curve represents the semiclassical result calculated by Mott formula. The red curve corresponds to the dissipationless limit. The onset of the EQL is marked by a vertical dashed line, and dashed lines labeled by equations describe different regimes in the dissipationless limit. }
    \label{fig:9}
\end{figure}

In Fig.~\ref{fig:9} we plot the Seebeck coefficient as a function of magnetic field  across the full range of different regimes of magnetic field.  In addition to the three semiclassical regimes given by Eqs.~(\ref{eqn:33})-(\ref{eqn:35}), the three regimes corresponding to the dissipationless limit, Eqs.~(\ref{eqn:37})-(\ref{eqn:39}), can also be seen. 
The semiclassical (blue curve) and dissipationless (red curve) calculations coincide in the interval $B_{\text{H}} \ll B \ll B_{\text{EQL}}$. 

\begin{figure}
    \centering
    \includegraphics[width=0.9\columnwidth]{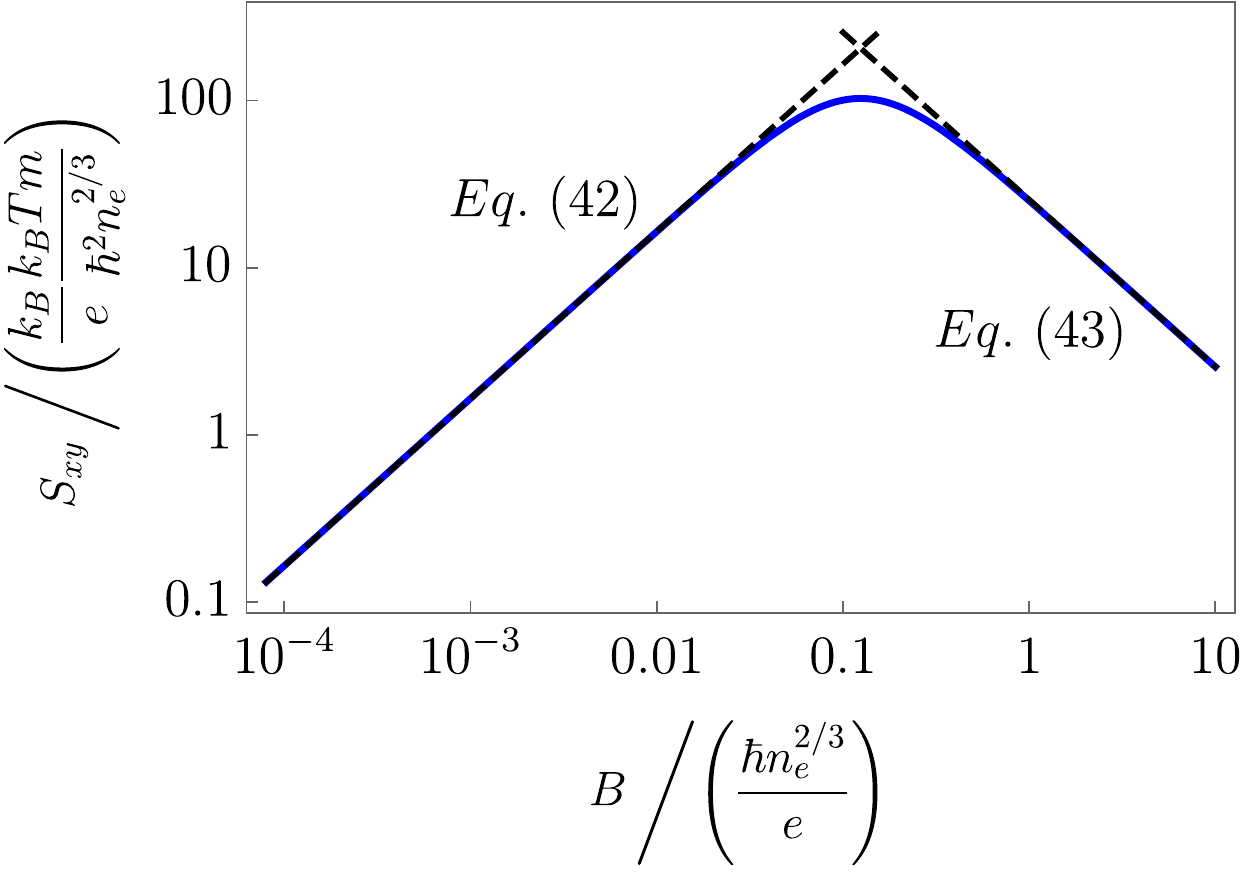}
    \caption{A plot of the Nernst coefficient $S_{xy}$ for an almost-completely compensated semiconductor. The material parameters are the same as in Figs.~\ref{fig:8} and \ref{fig:9}.}
    \label{fig:10}
\end{figure}

The Nernst coefficient can also be derived from the Mott formula in the semiclassical regime. This derivation gives
\begin{align}
S_{xy}\approx &\frac{\pi^{2/3}g^{2/3}}{3^{2/3}}\frac{k_B}{e}\frac{k_BT e\tau B}{\hbar^2n_e^{2/3}},&& B\ll B_H,\label{eqn:41}\\
S_{xy}\approx &\frac{4\pi^{2/3}g^{2/3}}{3^{2/3}}\frac{k_B}{e}\frac{k_BTm^2 n_e^{4/3}}{e\hbar^2\tau B (\Delta n)^2},&&  B\gg B_H.\label{eqn:42}
\end{align}
As in the Dirac/Weyl case, the value of $S_{xy}$ grows linearly with $B$ at $B \ll B_H$ and achieves a maximum of order $S_{xy} \sim (k_B/e) \times (k_BT/\ve_F) \times n_e/(\Delta n)$ at $B \sim B_H$.
This behavior is shown in Fig.~\ref{fig:10}.

\section{Conclusion}
\label{sec:concl}

In this paper we have presented a generic result for strong magnetic field enhancement of the thermopower in compensated conductors. 
The large magnitude of thermopower in these systems is somewhat surprising, since in the absence of magnetic field compensated systems have very small thermopower, owing to the near cancellation of the electron and hole contributions. In a magnetic field, however, electrons and holes have a component of their $\vec{E} \times \vec{B}$ drift motion that allows them to contribute additively to the heat current (see Sec.~\ref{sec:semi-quantitative}), and this large heat current drives the thermopower up.  

We emphasize that at sufficiently large magnetic field the thermopower is enhanced not just above their small zero-field value, but well above the value $S_{xx} \sim (k_B/e)\times(k_BT/\ve_F)$ associated with single-band conductors. This field enhancement begins as soon as the magnetic field becomes large enough that $\omega_c \tau \gg 1$, and it does not require the much more stringent conditions associated with large Hall angle or the extreme quantum limit, which are necessary in order to see magnetic field enhancement of thermopower in the single band case \cite{Skinnereaat2621, PhysRevB.99.155123}.

Our primary result, which is the $\sim B^2$ enhancement of thermopower beginning at $B > B_1$, requires the simultaneous existence of two strong inequalities. The first is $\omega_c \tau \gg 1$, which enables strong $\vec{E} \times \vec{B}$ drift of carriers that enhances the heat current $J^Q_x$. The second condition is $\sigma_{xy} \ll \sigma_{xx}$, which implies that the electrical resistance $\rho_{xx}$ grows quadratically with magnetic field and therefore the electrical current $J^e_x$ is reduced for a given applied voltage. These two conditions cannot be achieved simultaneously in single-band systems, and arise only because of the existence of a small parameter $(\Delta n) / n_e$. While our analysis has focused on the simplified case where both electron and hole bands have the same mobility, the generic material requirement for the existence of a regime $S_{xx} \propto B^2$ is
\be 
\Delta n \ll \left(\mu^e n_e + \mu^h n_h \right)  \frac{\min\{\mu^e,\mu^h\}}{\mu^e \mu^h} ,
\ee 
where $\mu^e$ and $\mu^h$ are the electron and hole mobilities, respectively.

We have also calculated the behavior of the thermopower within the EQL, using a picture based on quantum Hall edge states. We find that $S_{xx}$ behaves similarly to the results derived in Refs.~\cite{Skinnereaat2621, PhysRevB.99.155123}, except that it is enhanced by an overall factor $n_e/(\Delta n)$ that is very large when the degree of compensation is nearly complete.

The Nernst coefficient also exhibits an enhancement with increasing magnetic field, growing linearly with $B$ and attaining a large maximum value $S_{xy} \propto n_e/(\Delta n)$ at sufficiently large fields that $\sigma_{xy}$ is comparable to $\sigma_{xx}$.

We have not attempted to make a careful quantitative description of any particular experiment in this paper, but our results provide a potential explanation for the huge magnetothermoelectric effect observed in Ref.\ \cite{Han2020} in the compensated Weyl semimetal TaP. A calculation using no free parameters provides an estimate for $S_{xx}$ and $S_{xy}$ that is consistent both in trend and in order of magnitude with their results (Fig.~\ref{fig:7}). Our results may also provide an explanation for older experimental results on elemental bismuth \cite{PhysRevB.14.4381}, although a careful analysis remains to be done. 


Throughout this paper we have assumed that the EQL is achieved only at relatively large magnetic fields $B_{\text{EQL}} \gg B_{\text{H}}$, so that the Hall angle is large throughout the EQL.  In closing, let us briefly comment on the opposite case of $B_{\text{EQL}} \ll B_{\text{H}}$, for which $\Delta n$ is so small that $\sigma_{xy}$ is still small compared to $\sigma_{xx}$ at the onset of the EQL. In this case the EQL does not coincide with the ``dissipationless limit,'' and the thermopower in the regime $B_{\text{EQL}} \ll B \ll B_{\text{H}}$ must depend on the transport scattering rate. Describing current flow in this regime is difficult, since one cannot use the naive Boltzmann description (which is invalid inside the EQL) nor the description based on quantum Hall edge states (which does not account for scattering). If one nonetheless uses a naive Drude-type expression for the conductivity tensor in the regime $B_{\text{EQL}} \ll B \ll B_{\text{H}}$, together with the expression for $\alpha_{xy}$ in the EQL \cite{Zhang2020}, one arrives at a result $S_{xx} \sim (k_B/e) \times (e^3 k_B T v_F \tau^2 \Delta n) B^3/ (\hbar^4 n_e^{8/3})$. This result smoothly matches the semiclassical one [Eq.\ (\ref{eqn:25})] at $B \sim B_{\text{EQL}}$, as well as the expression for large Hall angle within the EQL [Eq.\ (\ref{eqn:29})] at $B \sim B_{H}$, so we suspect that it is broadly correct.  A more careful analysis remains to be done, however, in order to understand this regime.

Finally, let us comment on the constraints imposed by Onsager symmetry on the field-dependence of the Seebeck and Nernst coefficients. These symmetries demand that the value of the Seebeck coefficient is independent of the sign of the magnetic field (whether it points in the $+z$ or $-z$ direction), while the Nernst coefficient changes sign when $B$ is flipped. These dependencies are apparent in our semi-classical calculations
[Eqs.~(\ref{eqn:24})-(\ref{eqn:26}) and ~(\ref{eqn:33})-(\ref{eqn:35}) for $S_{xx}$ and Eqs.~(\ref{eqn:30})-(\ref{eqn:32}) and ~(\ref{eqn:41}),(\ref{eqn:42}) for $S_{xy}$]. In the dissipationless limit, however, one should be careful to note that the sign of the heat current $I_y^Q$ carried by edge states depends on the field direction. Our Eq.~(\ref{eqn:20}) assumes that $B$ points in the $+z$ direction; when the magnetic field is flipped, the sign of Eq.~(\ref{eqn:20}) is inverted. Inserting this sign correctly gives the invariance of $S_{xx}$ with field direction.

\ 

\acknowledgments 
The authors thank Joseph P.~Heremans, Nandini Trivedi and Dung Vu for their useful suggestions. This work was primarily supported by the Center for Emergent Materials, an NSF-funded MRSEC, under Grant No. DMR-2011876.

\bibliography{reference.bib}

\end{document}